\documentclass[twocolumn,showpacs,preprintnumbers,prl,amsmath,amssymb]{revtex4}
\usepackage{graphicx}% Include figure files
\usepackage{dcolumn}% Align table columns on decimal point
\usepackage{bm}% bold math
\usepackage{amsmath}

\begin{document}

\title{Tuning of the Fano Effect through a Quantum Dot in
an Aharonov-Bohm Interferometer}

\author{Kensuke Kobayashi, Hisashi Aikawa} 
\author{Shingo Katsumoto$^\dagger$}
\author{Yasuhiro Iye$^\dagger$}

\affiliation{ Institute for Solid State Physics, University of Tokyo,
5-1-5 Kashiwanoha, Chiba 277-8581, Japan\\ $^\dagger$Also at CREST,
Japan Science and Technology Corporation, Mejiro, Tokyo 171-0031,
Japan }

\date{\today}

\begin{abstract}
The Fano effect, which arises from an interference between a localized
state and the continuum, reveals a fundamental aspect of quantum
mechanics.  We have realized a tunable Fano system in a quantum dot
(QD) in an Aharonov-Bohm interferometer, which is the first convincing
demonstration of this effect in mesoscopic systems. With the aid of
the continuum, the localized state inside the QD acquires itinerancy
over the system even in the Coulomb blockade. Through tuning of the
parameters, which is an advantage of the present system, unique
properties of the Fano effect on the phase and coherence of electrons
have been revealed.
\end{abstract}
\pacs{73.21.La, 85.35.-p, 73.23.Hk, 72.15.Qm}

% 73.21.La    Quantum dots  
% 73.23.Hk    Coulomb blockade; single-electron tunneling  
% 85.35.-p   Nanoelectronic devices 
% 72.15.Qm Scattering mechanisms and Kondo effect (see also 75.20.Hr
%       Local moments in compounds and alloys; Kondo effect, valence
%       fluctuations, heavy fermions)

\maketitle

When a discrete energy level is embedded in a continuum energy state
and there is coupling between these two states, a resonant state
arises around the discrete level. In 1961, Fano
proposed~\cite{FanoPRB1961} that in such a system a transition from an
arbitrary initial state occurs through the two interfering
configurations --- one directly through the continuum and the other
through the resonance level --- and that this quantum mechanical
interference yields a characteristic asymmetric line shape in the
transition probability.  This is the Fano effect, a ubiquitous
phenomenon observed in a large variety of experiments including
neutron scattering~\cite{AdairPR1949}, atomic
photoionization~\cite{Fano1986}, Raman
scattering~\cite{CerdeiraPRB1973}, and optical
absorption~\cite{FaistNature1997}.  While a statistically averaged
nature of the system containing contributions from numerous sites is
observed in these experiments, the Fano effect is essentially a
single-impurity problem describing how a localized state embedded in
the continuum acquires itinerancy over the
system~\cite{MahanMPP}. Therefore, an experiment on a single site
would reveal this fundamental process in a more transparent way.
While the single-site Fano effect has been reported in the scanning
tunneling spectroscopy study of an atom on the
surface~\cite{MadhavanScience1998,LiPRL1998} or in transport through a
quantum dot (QD)~\cite{GoresPRB2000}, there is little, if any,
controllability in either case since the coupling between the discrete
level and the continuum is naturally formed.

In this Letter, we report the first tunable Fano experiment.  We have
clarified characteristic transport properties arising from this
effect, such as the delocalization of the discrete level and the
excitation spectra of the Fano system. External control of the
relative phase between a localized state and the continuum indicates
that the Fano parameter should be treated as a complex number.

\begin{figure}
\includegraphics[width=0.9\linewidth]{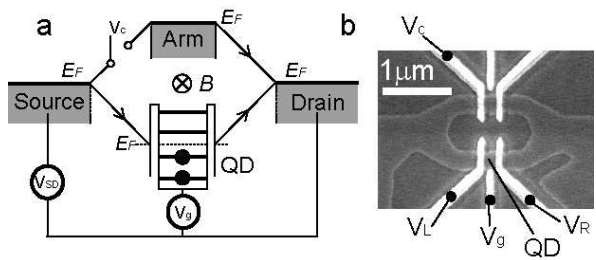}
\caption{\label{SampleFig} (a) Schematic representation of the
experimental setup.  An electron injected from the source traverses
the ring along two different paths through the continuum in the arm
and the discrete level inside the QD and interferes before the drain.
This corresponds to an artificial single-site Fano system.  (b)
Scanning electron micrograph of the correspondent device fabricated by
wet-etching the 2DEG at an AlGaAs/GaAs heterostructure.  The white
regions indicate the Au/Ti metallic gates.  The three gates ($V_R$,
$V_L$, and $V_g$) at the lower arm are used for controlling the QD and
the gate at the upper arm is for $V_C$. }
\end{figure}

To realize a well-defined Fano system, we designed an Aharonov-Bohm
(AB) ring with a QD embedded in one of its arms as seen in
Fig.~\ref{SampleFig}~(a), similar to those in previous
studies~\cite{YacobyPRL1995,SchusterNature1997,vanderWielScience2000,JiScience2000}.
The AB ring is essentially a double-slit interferometer of electrons.
In contrast, the QD~\cite{KastnerPT1993}, a small electron droplet
isolated from its leads by tunneling barriers, has discrete energy
levels arising from the electron confinement and the charging energy
that is much larger than the thermal energy $k_{B}T$ ($k_B$ is the
Boltzmann constant, and $T$ is the temperature). In the Coulomb
blockade (CB) regime, the single-particle level in the QD can be
controlled electro-statically by the gate voltage ($V_g$), and only
when the level matches the chemical potential of the leads conduction
through the QD is allowed. Thus, our ``modified'' AB interferometer
has the continuum energy state in one arm and the discrete energy
level in the QD in the other arm.  If the coherence of the electron is
fully maintained during tunneling through the QD and during
propagation through the arm, interference of traversing electrons will
occur through two different configurations.  Physically, this
situation is exactly a realization of the Fano
system~\cite{RyuPRB1998,KangPRB1999,WohlmanCM2001,KimCM2001}.  The
present system is unique in that it is controllable through several
parameters; the position of the discrete level inside the QD, the
coupling between the continuum and the discrete levels, and the phase
difference between two paths are controlled by $V_g$, the control gate
voltage ($V_C$), and the magnetic field ($B$) penetrating the ring,
respectively.

Figure~\ref{SampleFig}~(b) shows the fabricated Fano system on a
two-dimensional electron gas (2DEG) system at an AlGaAs/GaAs
heterostructure (mobility $=9\times 10^5$~cm$^2/$Vs, sheet carrier
density $=3.8\times10^{11}$~cm$^{-2}$, and electron mean free path
$l_e \sim 8$~$\mu$m).  The ring-shaped conductive region was formed by
wet-etching the 2DEG.  The length of one arm of the ring is $L \sim
2$~$\mu$m.  The Au/Ti metallic gates were deposited to control the
device. The three in the lower arm were used for defining a QD (with
$\sim$ 80 electrons contained in its area about
0.15$\times$0.15~$\mu$m$^2$) and the gate in the upper arm is $V_C$
for switching the transition through the continuum state on and
off. Measurements were performed between 30~mK and 800~mK in a
dilution refrigerator by a standard lock-in technique in the
two-terminal setup with an excitation voltage of 10~$\mu$V (80~Hz)
between the source and the drain.  Noise filters were inserted into
every lead at $T < 1$~K as well as at room temperature.

\begin{figure}
\includegraphics[width=0.90\linewidth]{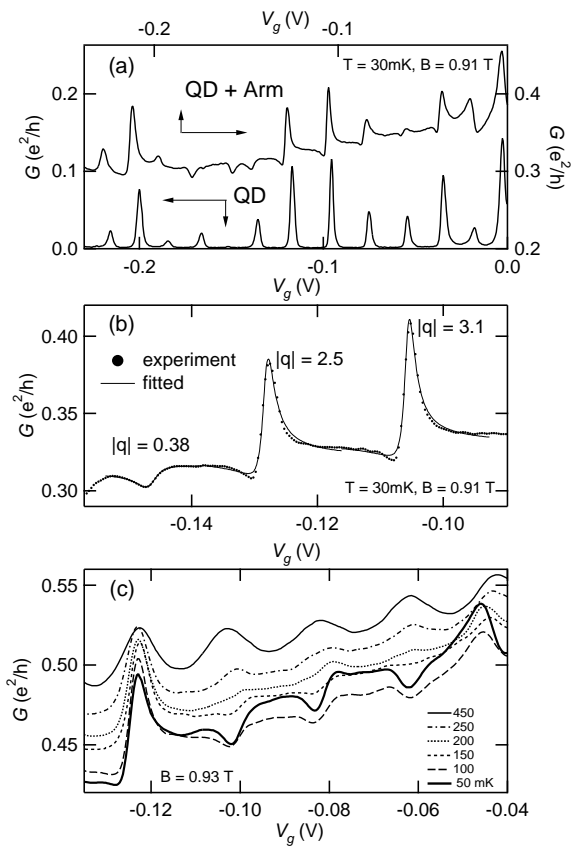}
\caption{\label{FanoOscillation} (a) Typical Coulomb oscillation at
$V_C = -0.12$~V with the arm pinched off, and asymmetric Coulomb
oscillation at $V_C = -0.086$~V with the arm transmissible. The latter
shows a clear Fano effect. Both of them were obtained at $T=30$~mK and
$B=0.91$~T. (b) The typical results of the fitting to the Fano
lineshape with $|q|=0.38$, 2.5, and 3.1. (c) The Fano effect measured
at several temperatures at $B=0.93$~T.  It gradually disappears as the
temperature increases. }
\end{figure}

First, we pinched off the upper arm by applying large negative voltage
on $V_C$. The QD was defined in the lower arm by tuning the side-gate
voltages ($V_L$ and $V_R$). The lower panel of
Fig.~\ref{FanoOscillation}~(a) shows the pronounced peaks in the
conductance $G$ through the QD as sweeping $V_g$, namely, a typical
Coulomb oscillation expected for QDs in the CB regime. The small
irregularity of the peak positions reflects that of the addition
energy and supports the occurrence of transport through each single
level inside the QD.

Next, we made the upper arm conductive. Because the control gate and
the QD are well separated electrostatically, a clear one-to-one
correspondence is observed between the two results in
Fig.~\ref{FanoOscillation}~(a), ensuring that the discreteness of the
energy levels in the QD is maintained. It is noteworthy that the line
shapes of the oscillation become very asymmetric and show even dip
structures. This is a clear sign of the Fano effect. Indeed, each peak
can be well fitted by the Fano line shape~\cite{FanoPRB1961}
$G(\epsilon)$ of the form
\begin{equation}
G(\epsilon) \propto \frac{(\epsilon+q)^2}{{\epsilon}^2+1},\quad
\epsilon=\frac{V_g-V_{0}}{\Gamma/2}
\label{FanoShape}
\end{equation}
where $V_{0}$ is the energy of the resonance position, $\Gamma$ is the
width of the resonance.  The real parameter $q$, which is the ratio of
the matrix elements linking the initial state to the discrete and
continuum parts of the final state, serves as a measure of the degree
of coupling between both.  The line shape analysis for several
resonance levels gives $|q|= 0.2$--$7$ with $\Gamma \sim 3$~meV as the
typical results of the fitting are shown in
Fig.~\ref{FanoOscillation}~(b).  The observed $|q|$ values are
correlated with the values of the conductance of the original Coulomb
peak in that a large peak tends to have a large $|q|$.  The dip
structure with $|q|< 1$ indicates a strong destructive interference,
supporting that the electron passing the QD retains sufficient
coherency to interfere with the one passing the arm in spite of the
significant charging effect inside and around the QD. This is in
contrast with the previous
reports~\cite{YacobyPRL1995,SchusterNature1997,vanderWielScience2000,JiScience2000},
where the coherence was limited to a fraction of the transmission
through the QD.

The Fano effect was found to be prominent within several specific
magnetic field ranges such as around $B \sim$~0.3, 0.9, and 1.2~T,
while it was less pronounced in the other ranges.  This implies that
the coherence of the transport through the QD strongly depends on $B$.
A similar role of $B$ is reported in the Kondo effect in a
QD~\cite{vanderWielScience2000}. Such phenomena arise from the change
of the electronic states caused by $B$ in each mesocopic system, while
this remains to be clarified theoretically.

When the temperature increased, decoherence increased and the
asymmetric Fano line shapes gradually evolved into an ordinary
Lorentzian line shape corresponding to $|q| \rightarrow \infty$ as
clearly seen in Fig~\ref{FanoOscillation}~(c).  Due to the loss of
coherence over the interferometer at $T \ge 450$~mK, the system is
simply a classical parallel circuit of the QD and the arm, being no
longer in the Fano state.

\begin{figure}
\includegraphics[width=0.8\linewidth]{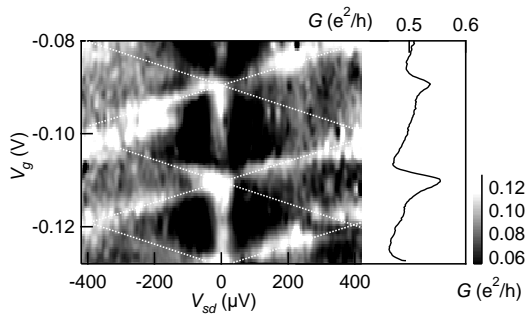}
\caption{\label{FanoVsd} Differential conductance obtained as a
function of $V_g$ at $T=30$~mK and $B=0.92$~T. The corresponding Fano
line shape is also shown in the right panel. The zero-bias conductance
peak exists in the CB region with a Coulomb diamond superimposed. The
edge of the CB region is emphasized with white dashed
lines. Incoherent contribution from the differential conductance of
the upper arm, which shows slight non-Ohmic behavior at finite
$V_{sd}$, has been subtracted from the data.}
\end{figure}

We measured the differential conductance at the lowest temperature as
a function of both source-drain bias ($V_{sd}$) and $V_g$.
Figure~\ref{FanoVsd} depicts that the resonating conductance peak of
the width $\sim 70$~$\mu$eV (colored white) stretches along the line
of $V_{sd}= 0$~V with the Coulomb diamond superimposed.  The
appearance of the zero-bias conductance peak even in the CB region
indicates that the transmission through the QD is now allowed due to
the aid of the continuum in the opposite arm.  Such delocalization of
the electron in the CB region is highly analogous with that observed
in the QD in the Kondo
regime~\cite{Goldhaber-GordonNature1998,CronenwettScience1998,SchmidPhysicaB1998},
although the mechanism is different; Delocalization in the Kondo dot
occurs through resonating spin singlet-pair formation, while resonance
due to the configuration interaction between the discrete state and
the continuum is the cause in the Fano regime~\cite{MahanMPP}.

\begin{figure}
\includegraphics[width=0.80\linewidth]{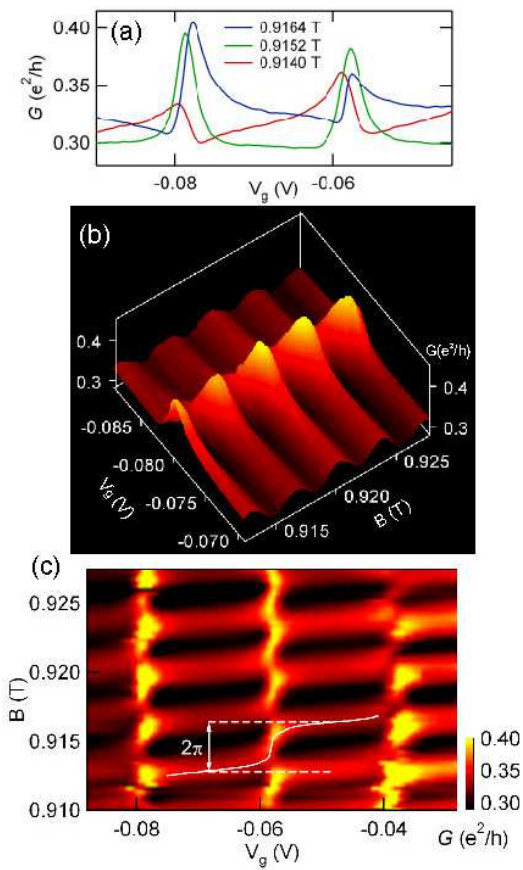}
\caption{\label{FanoAB} (a) Conductance of two Fano peaks at 30~mK at
the selected magnetic fields. The direction of the asymmetric tail
changes between $B=0.9140$ and 0.9164~T and the symmetric shape
appears in between.  (b) Conductance of one Fano peak as a function of
$V_g$ and $B$ at 30~mK. Note the large AB oscillation of the same
order as the resonance peak.  (c) Conductance of the Fano peaks as a
function of $V_g$ and $B$ at 30~mK.  AB oscillation exists even at the
midpoint of the resonances.  The white line represents the AB phase as
a function of $V_g$. Note that the AB phase changes by $2\pi$ through
the resonance, and all the resonances are in phase. }
\end{figure}

In our system, the discrete level and the continuum are spatially
separated, allowing us to control Fano interference via the magnetic
field piercing the ring as shown in Figs.~\ref{FanoAB}~(a), (b), and
(c).  The line shape changes periodically with the AB period of $\sim
3.8$~mT, which agrees with that expected from the ring
dimension. Figure~\ref{FanoAB}~(b) illustrates that the oscillation
amplitude at the conductance maximum is of the same order as the net
peak height, again ensuring that the transmission through the QD
occurs coherently. As $B$ is swept, an asymmetric line shape with
negative $q$ continuously changes to a symmetric one and then to an
asymmetric one with positive $q$; the sign of interference can be
controlled by the AB effect. Typical results are presented in
Fig.~\ref{FanoAB}~(a). The Fano effect is usually characterized by an
asymmetric line shape, while a perfect symmetric line shape is
obtained at a specific magnetic field here. Since the magnetic field
mainly affects the phase difference between the two paths through the
resonant state and the continuum, the aforementioned periodic behavior
is most likely explained systematically by introducing a complex
number $q$ whose argument is a function of $B$ or the AB flux,
although an expression of such $q$ applicable to our case is not known
at present. Here, Eqn.~(\ref{FanoShape}) is generalized to
$G(\epsilon) \propto |\epsilon+q|^2/({\epsilon}^2+1) =
\left((\epsilon+\Re q)^2+(\Im q)^2\right) /
({\epsilon}^2+1)$. Qualitatively, even when the coupling strength
$|q|$ is almost independent of $B$, the $B$-dependence of $\arg(q)$
yields asymmetric and symmetric line shapes of $G$ for $\Re q \gg \Im
q$ and $\Re q \ll \Im q$, respectively. In the original work by
Fano~\cite{FanoPRB1961} and most of the subsequent studies based on
his theory, the asymmetric parameter $q$ has been implicitly treated
as a real number, while this is valid only when the system has the
time-reversal symmetry and thus the matrix elements defining $q$ can
be taken as real. Our experiment indicates that when this condition is
broken, for example, by applying the magnetic field, $q$ should be a
complex number, which has not been explicitly recognized.

Figure~\ref{FanoAB}~(c) represents the result over three Fano
resonances in the $V_g$-$B$ plane. Clear AB oscillation is observed
even at the conductance valley between the resonant peaks. This
provides another evidence that the state in the QD becomes delocalized
with the aid of the continuum.  In Fig.~\ref{FanoAB}~(c) we also plot
the conductance maximum as a function of $V_g$, where the phase is
observed to change by $2\pi$ rapidly but continuously across the
resonance.  Since the measurement was performed in the two-terminal
setup that allows only phase changes by multiples of $\pi$ due to
reasons of symmetry, the continuous behavior of the AB phase is
unexpected, but may be attributed to the breaking of the time-reversal
symmetry~\cite{LeePRL1999,OudenaardenNature1998}.  The AB phase
changes only slightly at the conductance valley and, therefore, all
the adjacent Fano resonances are in phase, indicating that the
resonance peaks are correlated to each
other~\cite{LeePRL1999}. Several theoretical predictions on the
behavior of the AB phase in the Fano system have been
reported~\cite{RyuPRB1998,KangPRB1999,WohlmanCM2001,KimCM2001}, while
none of them perfectly reproduces the overall behavior discussed
above.  Further study is needed to fully understand the behavior of
the AB phase coexistent with the Fano effect.

It is also worth comparing our result with those in previous
two-terminal experiments for a normal QD~\cite{YacobyPRL1995} and a
Kondo QD~\cite{vanderWielScience2000}. In the former case, it was
found that the AB phase jumps by $\pi$ at the resonant peak and that
the adjacent Coulomb peaks are in phase. For a Kondo QD, no phase
change is observed in the Kondo valley and only one of the two Coulomb
peaks located at the side of the valley exhibits the $\pi$ phase flip.
Thus, the behavior of the AB phase in the present Fano system is
qualitatively different from both.  At this moment, the observed phase
of electrons through a QD, especially through a Kondo
QD~\cite{vanderWielScience2000,JiScience2000}, remains open, where the
Fano effect is recently predicted to play an important
role~\cite{WohlmanCM2001,BulkaPRL2001,HofstetterPRL2001}.  Our result
provides an important experimental indication that this effect
critically affects the phase evolution of electrons through such a
quantum system similar to ours, particularly when the coherence is
highly preserved.

In conclusion, we have clarified a peculiar quantum transport through
the QD-AB-ring hybrid system, which is caused by the Fano effect due
to the sufficiently coherent transport through the QD and the
ring. While this effect has been observed in a variety of physical
systems, the present system is the first convincing realization of a
Fano system that can be tuned. Delocalization of the discrete levels
in the QD due to this effect shows up both in the resonating zero-bias
peak in the differential conductance and in the considerable AB
amplitude in the CB. Controlling of the Fano line shape by the
magnetic field has revealed that the Fano parameter $q$ will be
extended to a complex number.  The behavior of the AB phase is found
to be very different from the previous results, indicating that our
system is not simply a QD with an adjective reference arm but should
be regarded as a novel quantum system.

This work is supported by a Grant-in-Aid for Scientific Research and
by a Grant-in-Aid for COE Research (``Quantum Dot and Its
Application'') from the Ministry of Education, Culture, Sports,
Science, and Technology of Japan.

\end{document}